\documentclass[aps,prl,twocolumn,float]{revtex4}
\usepackage{amsmath,bm,epsfig}
\let\*\cdot
\def\<{\left\langle} \def\>{\right\rangle} \def\({\left(} \def\){\right)}
 \let\~\widetilde \let\^\widehat 

\def\be{\begin{equation}}\def\ee{\end{equation}}
\def\bea{\begin{eqnarray}}\def\eea{\end{eqnarray}}
\def\bse{\begin{subequations}}\def\ese{\end{subequations}}
\newcommand{\BE}[1]{\begin{equation}\label{#1}}
\newcommand{\BEA}[1]{\begin{eqnarray}\label{#1}}
\newcommand{\BSE}[1]{\begin{subequations}\label{#1}}

\newcommand{\eq}[1]{(\ref{#1})}%%  requires \eq{label}
\newcommand{\Eq}[1]{Eq.~(\ref{#1})}%%  requires \eq{label}
\newcommand{\Eqs}[1]{Eqs.~(\ref{#1})}%%  requires \eq{label}
\usepackage{pstricks}
\usepackage{pst-node}
\usepackage[ansinew]{inputenc}
\usepackage{amssymb,amsmath}

\def\BSE{\begin{subequations}}\def\ESE{\end{subequations}}
\let \= \equiv

\def\a{\alpha}

\def\o{\omega}

\def\be{\begin{equation}}       \def\ba{\begin{array}}

\def\ee{\end{equation}}         \def\ea{\end{array}}

\def\bea {\begin{eqnarray}}      \def\eea {\end{eqnarray}}

\def\bean{\begin{eqnarray*}}    \def\eean{\end{eqnarray*}}

\def\<{\langle} \def\({\left(}  \def\>{\rangle} \def\){\right)}

\newtheorem{exi}{Example}

\begin{document}

\title{Cluster Dynamics of  Planetary Waves}
\author{Elena Kartashova$^{\dag\ddag}$ and Victor S. L'vov$^\dag$}
 \email{Lena@risc.uni-linz.ac.at, Victor.Lvov@weizmann.ac.il}
  \affiliation{$^\dag$ Department
of Chemical Physics, The Weizmann Institute of Science, Rehovot
76100, Israel \\
$^\ddag$ RISC, J.Kepler University, Linz 4040, Austria \\
$^*$Theoretical Department, Institute for Magnetism, National Ac. of Sci., Kiev, Ukraine}%%

   \begin{abstract}
The dynamics of nonlinear atmospheric planetary waves is determined
by a small number of independent wave clusters consisting of a few
connected resonant triads. We classified the different types of
connections between neighboring triads that determine the general
dynamics of a cluster.  Each connection type corresponds to
substantially different scenarios of energy flux among the modes.
The general approach can be applied directly to various mesoscopic
systems with 3-mode interactions, encountered in hydrodynamics,
astronomy, plasma physics, chemistry, medicine, etc.
\end{abstract}


\maketitle

 \noindent
{\bf 1. Introduction}. Planetary-scale motions in the ocean and
atmosphere are due to the shape and rotation of the Earth,  and play
a crucial   role in the problems of weather and climate prediction
\cite{ped}. Oceanic planetary waves affect the general large-scale
ocean circulation, can intensify the ocean currents such as the Gulf
Stream, as well as push them off their usual course. For example, a
planetary wave can push the Kuroshio Current northwards and affect
the weather in  North America~\cite{Kuroshio}. Atmospheric planetary
waves detach the masses of cold, or warm, air that become cyclones
and anticyclones
 and are responsible for day-to-day weather patterns
at mid-latitudes~\cite{smth}. Recently a novel model of
intra-seasonal oscillations in the Earth atmosphere has been
developed \cite{KL-06} in terms of isolated resonant triads  of
planetary waves (for wave numbers $m, \ell \le 21$). The complete
cluster structure depends on the spectral-domain size, both for
atmospheric \cite{KK-07} and oceanic \cite{all-08} planetary waves.
In particular, an  enlargement of at least some of the clusters is
possible, with growing of the spectral domain. In some cases this
yields the energy flux between previously isolated clusters. To
justify the basic model~\cite{KL-06} one has to understand
 whether or not existing clusters are capable to adopt external
energy via this mechanism. In \cite{KM07} the isomorphism
(one-to-one correspondence) between clusters of similar structure
and corresponding dynamical systems has been established. This
allows for the study of the dynamical behavior of similar clusters.
It is shown both in numerical simulations \cite{zak4,T07} and in
laboratory experiments \cite{DLN06} that the dynamics of mesoscopic
wave systems does not obey statistical description (wave kinetic
equations). It rather needs a special investigation. \par In this
Letter we show that the general dynamics of big clusters in
mesoscopic systems with 3-mode interactions  is determined by the
connection types between neighbor triads. In particularly, we
analyzed clusters of atmospheric planetary waves in the spectral
domain $m, \ell \le 1000$ which allowed us to justify  the model
suggested in \cite{KL-06}.

{\bf 2. Triad dynamics}. Consider  three planetary (Rossby) waves
with frequencies $\omega_1$, $\omega_2$ and $\omega_3$, which
satisfy the conditions
 of time and space synchronism:%%
\be \label{res}
\begin{cases}
 \omega_1+\omega_2=\omega_3,\\
 m_1+m_2=m_3,\\
 |\ell_1-\ell_2| \le \ell_3 \le \ell_1+\ell_2,\\
 m_j \le \ell_j, \ j=1,2,3,\\
\ell_i \neq  \ell_j, \ i \neq j, \ i,j=1,2,3,\\
 \ell_1+\ell_2+\ell_3 \ \ \mbox{is odd},
 \end{cases}
\ee%%
and $\omega \sim m/[\ell(\ell+1)]$. First three equations correspond
to three-wave resonance on a sphere while two  last equations
provide non-zero coupling coefficient in the corresponding dynamical
system (see \cite{KL-06} for more details).
 This is  the simplest possible cluster that is described by the dynamical
system%%
 \BE{3}  \dot{B}_1=   Z B_2^*B_3,\quad
\dot{B}_2=   Z B_1^* B_3, \quad \dot{B}_3= -  Z B_1 B_2. %
\ee %%
Here  $\dot B_j\= d B_j/ dt$,  $B_i=\a_i A_i$ is a time derivative
with $\a_i$ being explicit functions of the longitudinal wave
numbers $\ell_j$, $A_j$ are modes amplitudes, and $Z$ is the
interaction coefficient which is also some function of the wave
numbers.
 The equations \eq{3} are symmetric with respect to the exchange of two low-frequency modes $1\Leftrightarrow 2$.
The mode with the highest frequency (which in this paper will be
denoted by the subscript ``~$_3~$") is a special mode. The
system~\eq{3} has two independent conservation laws
 \BEA{MR}
 \begin{cases}
 I_{23}=|B_2 |^2 + |B_3|^2 =( E\, N_1-H ){N_{23} }/{N_1 N_2
 N_3}\,,\\
 I_{13}= |B_1 |^2 + |B_3|^2 =( E\, N_2-H ) {N_{13 }}/{N_1 N_2
 N_3}\,,\\
 I_{12}=I_{13}-I_{23}= |B_1 |^2 -|B_2 |^2\,,
 \end{cases}
\eea%%
which are linear combinations of the energy $E$ and enstrophy $H$,
defined as by%%
\be  \label{ints} E =  E_1+E_2+E_3\,,\quad  H = N_1E_1 + N_2E_2 +N_
3E_3\ .  \ee %%
Here $ E_j$ is the energy of the $j$-mode and
$N_j=\ell_j(\ell_j+1)$. The solutions of \Eqs{3} are Jacobian
elliptic functions, and whether or not its dynamics is periodic is
determined by the energy in
 the $\omega_3$-mode  (for details see~\cite{KL-06}).

To understand the dynamics of the energy flow within a cluster, the
first step would be to initiate a small amount of chosen modes and
to study afterwards the  energy exchange  within a cluster. Thus, we
begin with discussing  the evolution of the triad amplitudes with
special initial conditions, when only one mode is substantially
excited. If $B_1(t=0)\gg B_2(t=0)$ and $B_1(t=0)\gg B_3(t=0),$ then
$I_{23}(t=0)\gg I_{13}(t=0)$. The integrals of motion are
independent of time, therefore $I_{13} \gg I_{23}$ at each instant
of time and hence $|B_1(t)|^2\gg |B_2(t)|^2$. Moreover,
$|B_1(t)|^2\gg |B_3(t)|^2$ at every instant. Indeed, the assumption
$|B_1(t)|^2 \lesssim |B_3(t)|^2$ yields $I_{13} \simeq I_{23}$,
which is not the case. This means that the $\omega_1$-mode, being
the only substantially exited   at $t=0$ can not share its energy
with the two other modes in a triad. The same is true for the
$\omega_2$-mode. In this context we call the modes with frequencies
$\omega_1<\omega_3$
 and $\omega_2<\omega_3$ passive modes, or \emph{P-modes}.

The conservation laws~\eq{MR} cannot restrict the growth of the
P-modes from initial conditions when only $\omega_3$-mode is exited.
In this case the P-mode amplitudes will grow exponentially:
$|B_1(t)|\,,\ |B_2(t)|\propto \exp [\, |Z B_3(t=0)| t]$ until all
the modes will have comparable magnitudes of the amplitudes.
Therefore we call the $\omega_3$-mode an active mode, or
\emph{A-mode}. The A-mode, being initially excited, is capable to
share its energy with two P-modes within a triad.

 \noindent
{\bf 3. Connection types within a cluster}. An arbitrary cluster in
our wave system is a set of connected triads. A cluster consisting
of two triads that are connected via one common mode is called a
\emph{butterfly}; a cluster of three triads with one common mode is
called a\emph{ triple-star}. The general dynamics of a cluster
depends
 on the type of the connecting mode, which is common for the  neighbor triads.
 Correspondingly, we can distinguish   three types of butterflies (with  PP-, AP- and AA-connections),
  four triple-stars (with PPP-, PPA-, PAA- and AAA-connections), etc.   We begin with considering
   the butterfly and the triple-star dynamics; we then  discuss an actual dynamics
   of the
    more involved  but concrete topology of  connected triads in atmospheric planetary waves.

 A  \emph{PP-butterfly} consists  of two   triads  $\ a\ $ and $\ b,\ $
with wave amplitudes
 $\ B_{ja}, \ B_{jb},\ $ $j=1,2,3$,  that are connected {\it via} one common mode,  $\ B_{1a}=B_{1b},\ $,
 which is passive in both triads. The equations of motion for this system
 read
 \bea \label{PP}
\begin{cases}
\dot{B}_{1a}=  Z_a B_{2a}^*B_{3a} +   Z_b B_{2b}^*B_{3b},\ B_{1a}=B_{1b}, \,, \\
\dot{B}_{2a}=  Z_a B_{1a}^* B_{3a}\,,\quad    \dot{B}_{2b}=  Z_b B_{1a}^* B_{3b}\,, \\
\dot{B}_{3a}=  - Z_{a} B_{1a} B_{2a} \,,\    \dot{B}_{3b}=  - Z_{b} B_{1a} B_{2b}\ . \\
\end{cases}
 \eea%
An examination of \Eqs{PP}  shows that  they  have three integrals
of motion: \BEA{int}
\begin{cases} I_{23a}=|B_{2a} |^2 + |B_{3a}|^2 \,, \quad
 I_{23b}= |B_{2b} |^2 + |B_{3b}|^2 \,, \\
 I_{ab}= |B_{1a}|^2 + |B_{3a} |^2 + |B_{3b}|^2 \ .
 \end{cases}%%
  \eea
The first two,  $I_{23a}$ and $I_{23b}$, do not involve the common
mode $ B_{1a}=B_{1b}$,
 and are similar to integral $I_{23}$,  \Eq{MR},  for an isolated triad.
Similarly to the case of the evolution of a triad from the initial
conditions with an excited passive mode, the following conclusion
can be made. If at $t=0$  the amplitudes of one triad substantially
exceed
 two remaining amplitudes of the butterfly, that is, if
$|B_{1a}|, |B_{2a}|, |B_{3a}|\gg |B_{2a}|, |B_{3a}|$, then this
relation persists. In other words, in a PP-butterfly any of two
triads, $a$ or $b$, having initially very small amplitudes, will be
unable to adopt  energy from the second triad during its nonlinear
evolution.

An \emph{AP-butterfly} consists  of two   triads  $\ a\ $ and $\ b,\
$ with wave amplitudes that are
   connected   via  the common mode  $\ B_{3a}=B_{1b} $, which is active in one triad
   ($a$ for concreteness) and is passive in the second triad ($b$). In this case equations and integrals  of motion are:
 \bea
 \begin{cases}\label{AP}
 \dot{B}_{1a}=  Z_a B_{2a}^*B_{3a}\,, \quad  \dot{B}_{3b}=  - Z_{b} B_{3a} B_{2b}\,, \\
\dot{B}_{2a}=  Z_a B_{1a}^* B_{3a}\,,\quad    \dot{B}_{2b}=  Z_b B_{3a}^* B_{3b}\,, \\
\dot{B}_{3a}=  - Z_{a} B_{1a} B_{2a}  + Z_b B_{2b}^* B_{3b}\,,
 \end{cases}\\
 \begin{cases}
 I_{12a}=|B_{1a} |^2 - |B_{2a}|^2 \,, \quad
 I_{23b}= |B_{2b} |^2 + |B_{3b}|^2 \,, \\ \label{APintC}
   I_{ab}=|B_{1a}|^2+|B_{3a} |^2 + |B_{3b}|^2\ . \end{cases}
  \eea%
  The integrals $I_{12a}$ and $I_{23b}$,  do not involve a
  common mode $\ B_{3a}=B_{1b} $; they are similar to the corresponding integrals $I_{12}$ and $I_{23}$,  for the
  isolated triad.
 In the case, when the  triad $a$ is excited at $t=0$  much
 stronger then the $b$-triad (in which case $\ I_{12a} \gg I_{23b}$) then the smallness
 of the positively definite integral of motion $I_{23b}$ prevents the triad $b$  from
 adopting energy from the triad $a$ during all the evolution. The situation is different,
  when triad $b$ is excited initially  and $\ I_{23b} \gg I_{12a}$. In this case the initial
  energy of the triad $b$ can be easily shared with the triad $a$. The smallness of $I_{12a}$ only
  required that during the evolution $|B_{1a}|\approx |B_{2a}|$.

 For an \emph{AA-butterfly} with a common active mode in both triads ($ B_{3a}=B_{3b}$) we have:%
\bea \label{AA}
\begin{cases}
\dot{B}_{1a}=  Z_{a} B_{2a}^* B_{3a}\,,\    \dot{B}_{1b}=  - Z_{b}  B_{2b}^*B_{3a}\,, \\
\dot{B}_{2a}=  Z_a B_{1a}^* B_{3a}\,,\quad    \dot{B}_{2b}=  Z_b B_{1b}^* B_{3a}\,, \\
\dot{B}_{3a}= - Z_a B_{1a}B_{2a} -   Z_b B_{1b}B_{2b} \ . \\
\end{cases}\\
\begin{cases}I_{12a}=|B_{1a} |^2 - |B_{2a}|^2 \,, \quad
 I_{12b}= |B_{1b} |^2 - |B_{2b}|^2 \,, \\ \label{AAint}
   I_{ab}=|B_{1a}|^2+ |B_{3a} |^2 + |B_{3b}|^2\ .
\end{cases}
 \eea%
 Again, the integrals  $I_{12a}$ and $I_{12b}$  do not involve a common mode $\ B_{3a}=B_{1b} $
 and are similar to  $I_{12}$, \Eqs{MR}, for an isolated triad.
Simple analysis of these integrals of motion shows that the energy
that is initially hold in one of the triads will be dynamically
shared between both triads.

Finally we consider one of the triple-star clusters, e.g. the
APP-star, in which a common mode is active in the $a$-triad  and
passive in the $b$- and $c$-triad: $B_{3a}=B_{1b}=B_{1c}$. This
system has four integrals of motion: \BEA{APPint}
\begin{cases}
I_{12a}=|B_{1a}|^2- |B_{2a}|^2\,, \\ I_{23b}= |B_{2b}|^2 + |B_{3b}|^2\,, \quad I_{23c}= |B_{2c}|^2 + |B_{3c}|^2\,, \\
 I_{abc}= |B_{1a}|^2 +|B_{3a}|^2+ |B_{3b}|^2+ |B_{3c}|^2\ .
\end{cases}
\eea Similarly to butterflies, there exists one integral of motion
for each connected triad, that does not involve the common mode:
these are the integrals $I_{12a}$, $I_{23b}$, and $I_{23c}$, which
are the same as the corresponding integrals in the isolated triad.
The integrals $I_{23b}$ and $I_{23c}$ prevent the $b$- and
$c$-triads (which are connected via a P-mode) from adopting energy
from the initially excited $a$-triad. In those cases when the $b$-
and/or $c$-triads are initially exited, the $a$-triad can freely
adopt their energy via the connecting A-mode.

 Any triad that is
connected to a cluster (no matter how big the cluster would be) via
its passive mode cannot adopt energy from the cluster, if the triad
is not excited initially. On the other hand, a triad that is
connected to any cluster via an active mode can freely adopt energy
from the cluster during its nonlinear evolution.

 \noindent

 \noindent
\begin{table}
\begin{tabular}{|c|c|c|c|c| c||c|c|}
  \hline %%
  Clust. &   $\mathcal{N}_1$ & Modes $[m,\ell]$ & $\mathcal{N}_2$ & Connecting triads \\ \hline

\hline
$\Delta_1$& $1$ & [4,12] [5,14] [9,13]& $-$  & $-$\\
\hline
 $\Delta_2$ & $2$ & [3,14] [1,20] [4,15] & 2.1 & [4,15] [10,24] [14,20]\\

   &  &  & 2.2 & [1,20] [14,29] [15,28]\\
   &  &  & 2.3 & [1,20] [15,75] [16,56]\\
\hline
 $\Delta_3$& $3$ & [6,18] [7,20] [13,19] &3.1 & [2,15] [5,24] [7,20] \\
\hline
 $\Delta_4$& $4$ & [1,14][11,21][12,20]& 4.1 & [1,14] [9,27] [10,24] \\
\hline \hline
$\bowtie_{5,6}$  & $ 5$ & [2,6] [3,8] [5,7] &  5.1& [4,14] [9,27] [13,20]    \\
    & $ 6$& [2,6] [4,14] [6,9]&  & \\
\hline
  &$ 7$ & [6,14] [2,20] [8,15] & 7.1& [2,20] [11,44] [13,35] \\
    $\bowtie_{7,8}$ & $ 8$ & [3,6] [6,14] [9,9] & 7.2 & [2,20] [30,75] [32,56]\\
   &  &  & 7.3 & [32,56] [26,114] [58,69]\\
\hline  $\bowtie_{9,10 } $  & $ 9$ & [3,10] [5,21] [8,14] & $-$ & $-$ \\
   &$ {10}$  & [8,11] [5,21] [13,13] & & \\

\hline \hline
  & $ {11} $ & {  [2,14]}[17,20][19,19]&11.1& [2,14] [18,27] [20,24]\\
   & $ {12}$ & {  [1,6]}  {  [2,14] [3,9]}&11.2& [6,44] [14,21] [20,24] \\

 & $ {13}$& {  [3,9] } [8,20] [11,14] &11.3 & [9,35] [11,20] [20,24] \\

  $\boxtimes_{11\!-\!16} $  & $ {14}$ & {  [1,6] [11,20] [12,15]} & 11.4& [3,20] [45,75] [48,56]\\

  & $ {15}$ & [9,14] [3,20]{   [12,15] }&  & \\

& $ {16}$ & [2,7] {  [11,20]} [13,14]&  & \\

\hline
\end{tabular}
\caption{In the first 3 columns  the following data
in the domain $m, \ell \le 21 $  are given: cluster's form, triad
numbers and modes within a cluster; in the last two -- numbers
of connecting triads and their modes are given that enlarge
corresponding cluster when spectral domain $m, \ell \le 1000 $ is
regarded.}
\end{table}%
\noindent {\bf 4.~Topology~and cluster dynamics  for atmospheric
planetary waves}. The frequencies of atmospheric planetary  waves
are%%
\BE{freq} \omega_ j= -2\,\Omega \, m_j/\ell_j (\ell_j+1)\ .\ee %%
The negative sign indicates wave propagation opposite to the
rotation of the Earth (east to west), opposite to the Earth rotation
(with frequency $\Omega$). The integers $\ell_j$ with  and $m_j\le
\ell_j$ describe the eigen-mode structure, which is $j$-spherical
harmonics with $\ell_j$ and $\ell_j-m_j$ zeros in the longitudinal
and latitudinal directions, correspondingly. The Diophantine
  \Eqs{freq} and $\omega_1+\omega_2=\omega_3$,  have many solutions, each of them
describing an exact resonance of \emph{ideal} planetary waves
(ignoring the real Earth topography, etc.). In this approximation we
can describe all important resonant triads in the Earth atmosphere,
see Table 1. First we restricted ourself to the so-called
meteorologically significant spectral domain with wave numbers
$\ell, m \le 21$~\cite{KL-06}.  In this domain we have found
in~\cite{KL-06} four isolated triads, denoted as $\Delta_1\dots
\Delta_4$, three PP-butterflies $\bowtie_{5,6}$,  $\bowtie_{7,8}$,
and $\bowtie_{9,10}$, involving of six  triads $\Delta_5\dots
\Delta_{10}$ and one, further  called \emph{caterpillar},
$\boxtimes_{11-16} $, consisting of six triads
$\Delta_{11}\dots\Delta_{16}$ with three PP-, one AP- and one
AA-connection.

\begin{figure}
\begin{center}
\includegraphics[width=3.5cm,height=2cm]{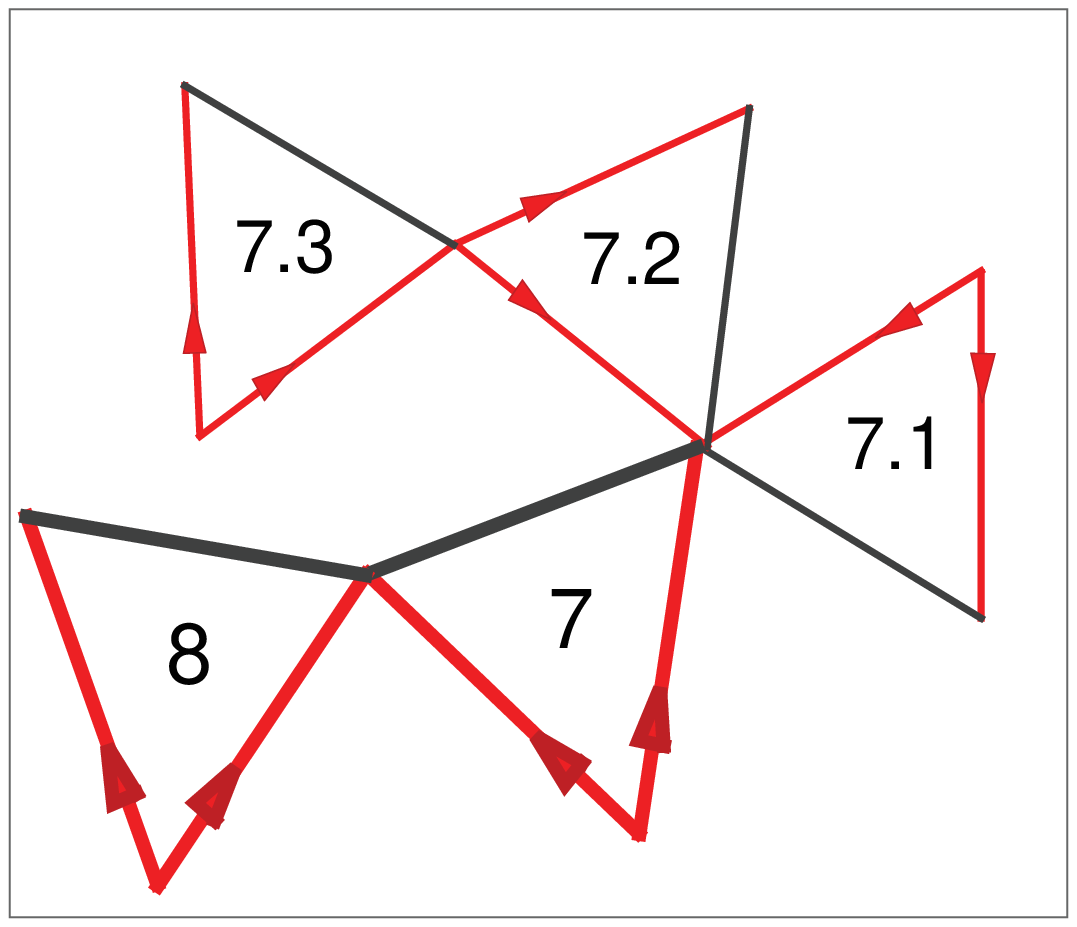}
\includegraphics[width= 5cm,height=3cm]{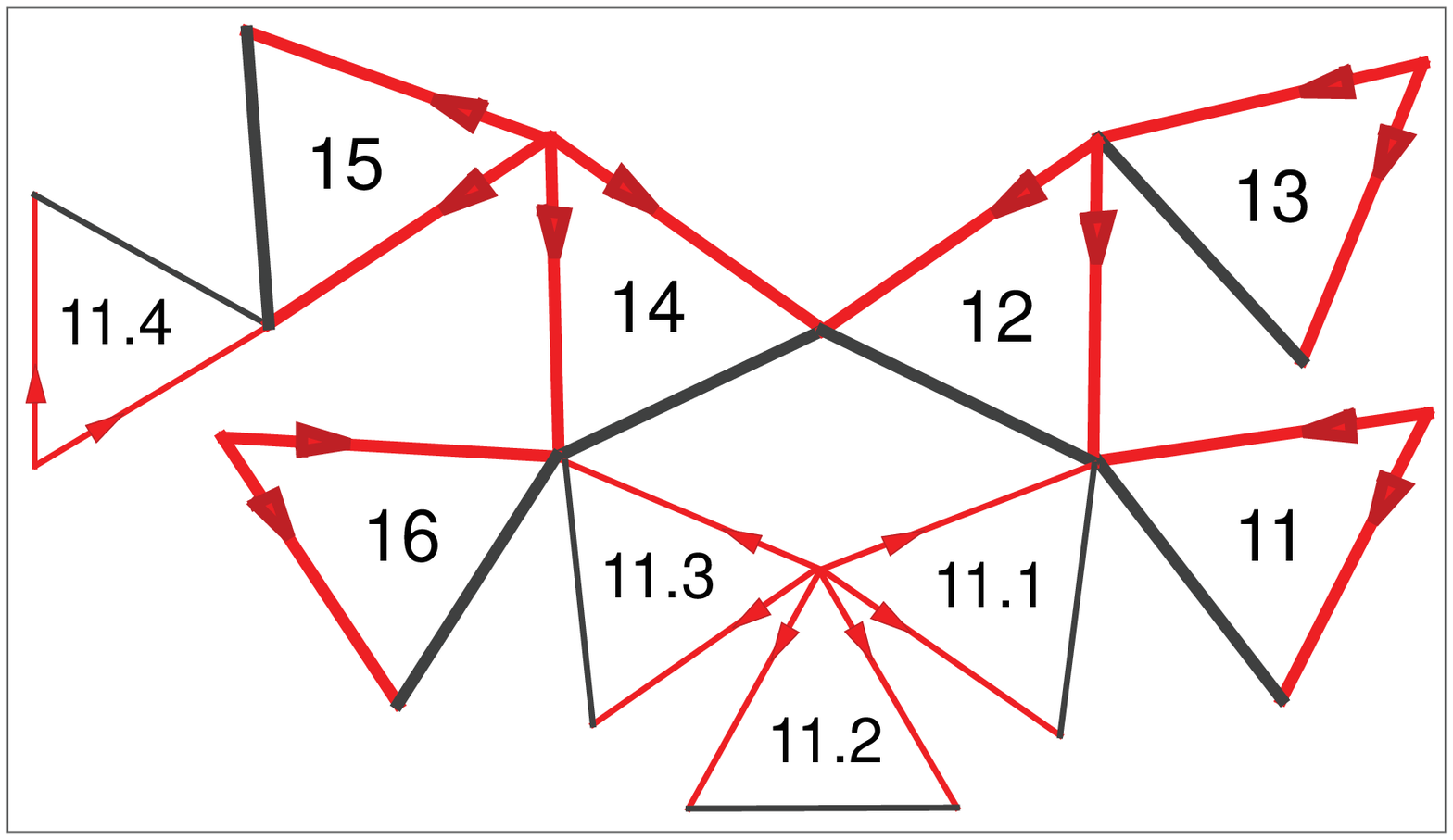}
\end{center}
\caption{\label{f:butterfly} Color online. Triads belonging to
butterfly $\bowtie_{2}$ (left) and to caterpillar (right)
 are drawn by bold (red) lines while new, connected to them  triads
appearing in spectral domain $m,\ell \le 1000$ are drawn by thin
lines. (Red) arrows are coming from an active mode and show
directions of the energy flux. The numbers inside each triangle
correspond to the numeration in Table 1.}
\end{figure}

In this Letter we show that the topological structure of clusters in
the
 extended  spectral domain $m, \ell \le 1000 $ is richer,
 in particular, some clusters are enlarged by new
 resonances formed by modes with  $m, \ell > 21. $ To illustrate
 this mechanism, we introduced following rows in the Table I: 1)
 $\mathcal{N}_1$ - the number of clusters in the spectral domain $m,
 \ell \le 21,$ with the same numeration as given in \cite{KL-06}; 2)
 "Modes" - resonance clusters belonging in the same spectral domain as
 in 1); 3) $\mathcal{N}_2$ - number of the additional clusters in
 the spectral domain $m, \ell \le 75;$ 4) "Connecting triads" -
 additional resonance clusters which appeared in the spectral domain
 as in 3). The topological structure of resonance clusters in the
 spectral domain $m, \ell \le 75$ is shown in Fig.\ref{f:butterfly}, where
 numbers inside of triangles correspond to the numeration
 $\mathcal{N}_1, \ \mathcal{N}_2$ in the Table 1.

\begin{figure}
\begin{center}
\includegraphics[width=8cm,height=4cm]{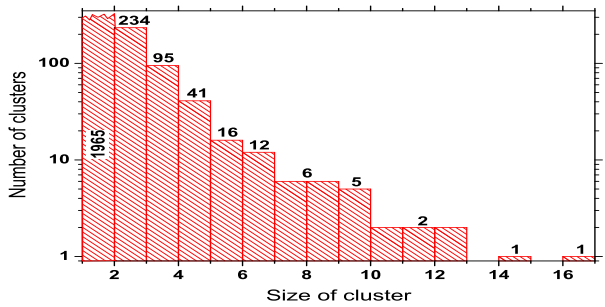}
\end{center}
\caption{\label{f:histogram}Color online.  Horizontal axes denotes
the number of triads in the cluster while vertical axes shows the
number of corresponding clusters.}
\end{figure}

There exist altogether 1965 isolated triads and 424 clusters
consisting of 2  to 3691 connected triads, among  them 235
butterflies, 95 triple-triad clusters, etc. - see the histogram in
Fig. \ref{f:histogram}. The three largest clusters consist of 14, 16
and 3691 connected triads.  For the clarity of presentation we did
not display on the histogram the largest  3691-cluster, which we
will call further the \emph{monster}.

It can be seen that $82.2\%$ of all clusters are
 isolated triads.  Their dynamics   has been investigated in \cite{KL-06} in all
 details. The
energy  oscillates between three modes in the triads, the period of
this oscillation being much larger than the wave period. It was
found to be inversely proportional
 to the root-mean-square   of the wave amplitude.

 235 clusters ( $\simeq 10.5 \%$) are butterflies. Among them there are
  131 PP-, 69 AP-  and 35 AA-butterflies. The butterfly dynamics is restricted
  by three integrals of motion, and it can be shown that the phase space of butterflies
   is four-dimensional. In \cite{KL-06} only PP-butterflies have
   been considered. For the initial conditions studied in this Letter their
 dynamics
  is  similar to that of two isolated triads.

Preliminary numerical simulations show that only in the case when
the initial levels of excitation in both triads are
  compatible, a periodic energy exchange is observed {\it not only}
  within  a triad but also between two connected triads.

 The 95 triple-triad clusters include  66 linear clusters with two pair connections, 25
 "3-stars" with triple connections and 4 -triangles (with three
pair connections).   A similar classifications can be performed for
all the remaining clusters.  For example, the monster includes one
mode (218,545), participating in 10 triads, three modes,
participating in 9 triads,  5 modes -- in 8 triads,  23 -- in 7, 50
-- in 6, 90 -- in 5, 236 -- in 4, 550 -- in 3, and 1428 modes -- in
2 triads (butterflies). The analysis of their dynamical behavior
depends critically on the type of connection, as was shown above.

For example, the 16-triad cluster can be divided into "almost
separated" parts (connected through PP-connections) parts:  5
triads, one AA- and one AP-butterfly, one AAP-star and one AAA-star
with an AP-connected triad. The overall qualitative conclusion is
that even big clusters are dynamically not very different from
separated small clusters with active connections, AA-butterflies,
AAA-stars, etc.

To clarify the dynamics of the first sixteen triads $\Delta_1\dots
\Delta_{16}$, it is  important to establish their (possible)
connection to the clusters in a bigger spectral domain. We have
found that the triad $\Delta_1$ remains isolated, $\Delta_3$ turned
into an isolated PA-butterfly. It can be proven~\cite{AMS} that
these objects remain  "forever" isolated, even if the size of the
bigger domains goes to infinity. The triads  $\Delta_2$ and
$\Delta_4$ became parts of the monster but are connected with it via
P-modes; in this sense they are practically separated. Three initial
butterflies, $\bowtie_{5,6},\bowtie_{7,8},\bowtie_{9,10}$ have
PP-connections and thus their dynamics does not differ much from the
dynamics of an isolated triad. Moreover, an increase  of the
spectral domain to $m, \ell \le 1000 $ does not change the situation
substantially: the butterfly $\bowtie_{9,10}$ remains isolated in an
arbitrarily large (even infinite) domain, $\bowtie_{7,8}$ became
part of a 5-triads cluster, also with PP- and PPP-type of
connections, (see Fig.\ref{f:butterfly}), $\bowtie_{5,6}$ is now
part of the monster but only via PP-connections.  The caterpillar
 gains one P-connected triad (11.4 in the Tab.~1)
and one AAA-star (11.1,\ 11.2 and 11.3 triads) with two
P-connections, see Fig.\ref{f:butterfly}. Therefore  almost all
16-triads, (except of the AA-butterfly $\bowtie_{15,16}$) can be
considered as completely  or as almost separated from the rest of
the system.

\noindent
 {\bf Conclusions}:

\textbullet~In  the physically relevant domain of atmospheric
planetary waves ($m\,,\ \ell \le 1000$, when the mode-scale is
larger than the height of the Earth atmosphere) we have determined
and described topology of all clusters that are formed by resonantly
interacting planetary modes. The cluster set contains
 isolated triads and sets of 2-, 3-, $\dots$, 16 and  3691 connected triads, with 2- 3-, $\dots$, 9-mode and (maximum)
 10-mode connections.

\textbullet~Analyzing the integrals of motion we suggested a
classification i) of triad modes into two types  - active (A) and
passive (P),  and ii) of connection types between triads - AA, AP and
PP. We have shown  that in AA-butterflies
 the energy can flow in both directions, in AP-butterflies  only from one triad to the
 second one (and not vice versa), while in PP-butterflies the triads are "almost isolated".
 We have also shown that the dynamical behavior of bigger clusters can
 be similarly characterized  by  connection types like
 AAA, AAP,..., etc.

\textbullet~As a first approximation, almost all triads in the
meteorologically significant domain of Tabl.~1,  can be considered
as completely or almost separated from the rest of atmospheric
planetary waves, and therefore energy oscillations within them can
lead to intra-seasonal oscillations in the Earth's atmosphere, as
suggested in~\cite{KL-06}.

\textbullet~ Our analysis is based on the structure of the dynamical
system~\eq{3}  and has two advantages: i) it   does not need to
exploit explicit form of the interaction coefficients $Z$; ii) it is
completely analytical without using numerics (which will be required
for more general initial conditions). Thus   our results can be used
directly for arbitrary resonant 3-wave systems governed by these
equations, e.g. drift waves, gravity-capillary waves, etc.

\textbullet~  A touchstone of any theory is, of course, an
 experiment. Numerical experiments with resonance clusters described
 by barotropic vorticity equation are now on the way, preliminary
 results confirm our theoretical conclusions. The next step of utmost
 importance would be to study some physical mechanisms that might
 destroy clusters. Speaking mathematically, one can always introduce
 big enough resonance width
 $$ \Omega = \o_1+\o_2-\o_3 > 0$$
 such that a substantial part of cluster's energy will be
 re-distributed among other waves {\it via} non-resonant interactions
 (see Fig.2, \cite{PRL}). From the physical point of view, the source of
 the resonance broadening might have substantially different reasons
 - from baroclinic instabilities due to the effects of the free
 surface at large scales to the effects due to the Earth topography at
 smaller scales to the increasing the level of turbulence, say, in
 summer due to increasing sun activity, and thus going into the
 regime of fully developed wave turbulence, to the inclusion of
 dissipation and forcing. This effects can also be combined, of
 course. The problem of the utmost importance is therefore to study
 the resonance clusters behavior in the situation when at least some
 of these effects are included. The analysis applied to climatic
 variations on geological scales "typically give indications of low
 dimensionality and (...) the hope of justifying the modelling of
 weather or climate in terms of a small set of ordinary differential
 equations" \cite{Ray}. But it does not mean, of course, that the
 overall energy flux will stay nicely regular. Indeed, as it was
 shown in \cite{Ray}, a special choice of instabilities included into
 \eq{3} will cause appearance of strange attractors. The study of
 this transition from regular to chaotic regimes in resonant clusters
 and mutual interrelations of relevant physical parameters is the
 subject of our further research.

 \textbullet~ Last not least. Whereas it is quite difficult to check
 experimentally the theory for planetary waves, it can be done much
 easier within a framework of laboratory experiments with some water
 waves, e.g. gravity-capillary waves. Some preliminary program of
 laboratory experiments with this type of waves has been worked out
 in \cite{GrCap1995} but only for simplest triad's clusters for at
 that time the algorithms of cluster computing \cite{KK06-1,KK06-2}
 were not developed yet.) Presently this program can easily be
 elaborated for clusters of some more complicated structure.
 Considerably more sophisticated preliminary work is needed for
 planning  laboratory experiments with gravity water waves. As it
 was shown in \cite{K07}, a 3-wave resonance system differs principally
 from wave systems in which 4- and more wave resonances are
 allowed. Indeed, in a 3-wave resonances system, each nonlinear
 resonance generates new scale, while already in a 4-wave system this
 is not  necessarily true. Indeed, beginning with 4-wave resonances,
 different types of energy fluxes have been pointed out:
 scale-resonances (as in 3-wave  system), angle-resonances (formed by
 two couples of wavevectors with pairwise equal lengths) and mixed
 cascades. In this case, clusters have more complicated structures
 presented explicitly in \cite{KNR08} for 4-wave interactions of
 gravity water waves. Qualitative dynamics of small clusters formed
 by resonant quartets is briefly as follows. "One wave mode may
 typically participate in many angle resonances and only one scale
 resonance. Thus, one can split large clusters into "reservoirs",
 each formed by a large number of angle quartets in quasi-thermal
 equilibrium, and which are connected with each other by sparse links
 formed by scale quartets."(see \cite{KNR08}). This means, in
 particular, that scale-resonances cause spectrum anisotropy. On
 the other hand, one can easily compute what ratio aspect of the
 sides  of a laboratory tank should be chosen in order to suppress
 scale-resonances. In this case, regular patterns on the water
 surface are to be expected, similar to what was observed in
 \cite{CHOS06,HPS06, SH07}. It would be interesting to study these
 experimental data in order to establish these nearly permanent
 patterns observed, which can be attributed to some specific resonance
 clusters. Existence of
  independent resonance clusters can shed some light on the origin of
 Benjamin-Feir instability \cite{BF67} or McLean instability
 \cite{McL82} (see also Discussion in \cite{KNR08}).

 Another interesting and even more complicated area of further
 investigations would be the study of the dynamical behavior of
 resonant quintets, sextets and so on. Methods developed in
 \cite{KK06-1, KK06-2} allow to compute the  clusters and
 corresponding wave frequencies. This information can be afterwards
 used, for instance, for investigating of some special types of
 shallow water instabilities \cite{FK05}. In systems
 containing simultaneously quartets and quintets \cite{Badulin95},
 the explicit construction of clusters can possibly yield the
 explanation of the competing regimes between these two types of
 instabilities. We point out here three possible scenarios due to the
 change of 1) boundary conditions, 2) frequency range under the
 study, and 3) initial distribution of energies among the modes of a
 cluster. In the first two cases, some quartets can be
 suppressed, thus turning the quintets into the principal clusters
 (and {\it vice versa}, of course). In the third case, a lot of
 different sub-scenarios are possible: quartet and quintet connected
 {\it via} one common mode can become "independent" if energy in the
 common mode is too small; a quartet can "die out" if its initial
 energy is not enough for nonlinear interactions; same for the
 quintet; energy can flux from the quartet to the quintet {\it via}
 the common mode, in dependence on the energy distribution among the
 other modes in the quartet thus yielding the dying out of the
 quartet; same for the quintet; etc. Combining the methods
 \cite{KK06-1,KK06-2} and results of \cite{Badulin95} will help to
 single out some of the possibilities immediately. For instance, one
 can compute wavenumbers corresponding to a resonant cluster, put
 them into the Eq.(2.11) from \cite{Badulin95} and check whether
 coupling coefficient(s) in this cluster is (are) non-zero. Another
 advantage would be construction of the topological structure of
 the complete cluster set as it is done in \cite{KNR08} instead of
 resonance curves: it gives a general overview of the nonlinear
 dynamics rather than a (possible) subset of resonances formed with a
 fixed wavevector (in particular cases this subset can be empty, of
 course). One more advantage would be the following. As it was shown in
 \cite{Badulin95}, instability analysis is based on the
 properties of the roots of the polynomial given by Eq.(3.10) (see
 also Fig.3a-3d therein). In some cases a qualitative  instability
 analysis can help but "to get qualitative results in general case
 one should solve (3.10)" (\cite{Badulin95}, p.313). This
 computations are very involved while the polynomial has degree 4 and
 its  coefficients depend on the dynamical coupling coefficients. On
 the other hand, knowledge of explicit wavenumbers for modes forming
 a resonance cluster will turn the coefficients of (3.10) into
 known constants and the problem can be easily solved.

 \noindent
{\bf Acknowledgements}. Authors express a special gratitude  to Ray
Pierrehumbert and Oleksii Rudenko for fruitful and stimulating
discussions, and to Rudolf Treumann for his valuable help during the
preparation of the final version of this paper. We acknowledge the
supports of the Austrian Science Foundation (FWF) under project
P20164-N18 "Discrete resonances in nonlinear wave systems" and  of
the Transnational Access Programme at RISC-Linz, funded by European
Commission Framework 6 Programme for Integrated Infrastructures
Initiatives under the project SCIEnce (Contract No. 026133). This
research was supported in part by the National Science Foundation
under Grant No. PHY05-51164. Authors are genuinely grateful to both
anonymous Referees whose suggestions made the form our paper more
clear and allowed to put our paper better in the context of physics
and meteorology. Authors are particularly obliged to the Referee B
who attracted our attention to the paper \cite{Badulin95}.

 \end{document}